\documentclass[a4paper,11pt]{article}
\usepackage{pos}

\usepackage[justification=RaggedRight]{subcaption}
\usepackage{physics}
\newcommand{\bonn}{
    Helmholtz-Institut f\"{u}r Strahlen- und Kernphysik,
    Rheinische Friedrich-Wilhelms-Universit\"{a}t Bonn, 53115 Bonn, Germany
}

\newcommand{\ias}{
    Institute for Advanced Simulation,
    Forschungszentrum J\"{u}lich, 54245 J\"{u}lich, Germany
}

\newcommand{\jsc}{
    JARA \& J\"{u}lich Supercomputing Center,
    Forschungszentrum J\"{u}lich, 54245 J\"{u}lich, Germany
}

\newcommand{\liverpool}{
    Department of Mathematical Sciences,
    University of Liverpool, Liverpool, L69 7ZL, United Kingdom
}

\newcommand{\casa}{
    Center for Advanced Simulation and Analytics (CASA),
    Forschungszentrum Jülich, 52425 J\"{u}lich, Germany
}

\title{Reducing the Sign Problem with simple Contour Deformation}
\author*[a, b, c]{Christoph G\"antgen} 
\author[a, b, d]{Evan Berkowitz}
\author[a, c,]{Thomas Luu} 
\author[e]{Johann~Ostmeyer}
\author[a, b, c, d]{Marcel Rodekamp}

\affiliation[a]{\ias}
\affiliation[b]{\casa}
\affiliation[c]{\bonn}
\affiliation[d]{\jsc}
\affiliation[e]{\liverpool}

\emailAdd{c.gaentgen@fz-juelich.de}
\emailAdd{e.berkowitz@fz-juelich.de}
\emailAdd{t.luu@fz-juelich.de}
\emailAdd{m.rodekamp@fz-juelich.de}
\emailAdd{j.ostmeyer@liverpool.ac.uk}

\abstract{We apply constant imaginary offsets to the path integral for a reduction of the sign problem in the
Hubbard model. These simple transformations enhance the quality of results from HMC
calculations without compromising the speed of the algorithm. This method enables us to efficiently
calculate systems that are otherwise inaccessible due to a severe sign problem. To support this claim,
we present observables of the C20 and C60 fullerenes. Furthermore, we demonstrate that at a certain
offset, the sign problem is completely lifted in the limit of large chemical potential.}

\begin{document}
\maketitle

\section{Introduction}
The Hubbard model named after John Hubbard originates from solid state physics. It is particularly well known for its ability to describe the transition from conducting to insulating metals. It describes a physical system as a spatial lattice with particles occupying the sites. These particles can \emph{hop} from one site to another, which is described by a kinetic term and usually restricted to neighboring sites. Multiple particles sharing the same site interact with each other via a localized potential that can be repulsive or attractive. Since the system is fermionic the Pauli exclusion principle holds. Optionally a chemical potential can be added to influence the number of charged particles within the system.
Here we only look at the fermionic Hubbard model with repelling potential which describes electrons on a lattice of ions
\begin{equation}\label{eq:HubbardHamiltonian}
  H = -\sum_{ x,y}{\kappa}_{x,y}\left(a^\dagger_{x\uparrow}a_{y\uparrow}^{} + a^\dagger_{x\downarrow}a_{y\downarrow}^{}\right) - \frac{{U}}{2}\sum_{x}{\left(n_{x\uparrow}-n_{x\downarrow}\right)}^2 - {\mu} \sum_x (n_{x\uparrow} + n_{x\downarrow})\,.
\end{equation}
Here $\kappa$ is the hopping matrix describing the hopping amplitudes, $U$ is the interaction potential and $\mu$ the chemical potential \cite{Hubbard1959}.
By applying a Hubbard-Stratonovich transformation on \autoref{eq:HubbardHamiltonian} we can formulate an action %\autoref{eq:HubbardAction} 
with continuous auxiliary fields $\phi$,
\begin{equation}\label{eq:HubbardAction}
  S=\sum_{x,t}\frac{\phi_{x,t}^2}{2\tilde{U}}-\log\det(M[\phi,\tilde{\kappa},\tilde{\mu}]M[-\phi,-\tilde{\kappa},-\tilde{\mu}]) \in \mathbb{C}
\end{equation}
which allow lattice field calculations \cite{Stratonovich}. 
A tilde indicates parameters in lattice units, e.g.\ $\tilde{U}=U \delta$ where $\delta = \beta/N_t$ is the lattice spacing.
The fermionic properties are encoded in the fermion matrices $M$, one for spin up  and one for spin down (or equivalently, one for particles and one for holes).  At non-zero $\mu$ there is no ergodicity problem with the \emph{exponential} discretization that we use~\cite{Wynen2019}. 

The term involving these matrices in~\autoref{eq:HubbardAction} can be complex for non-zero chemical potential or non-bipartite lattices.
This property makes the Hubbard model susceptible to the \emph{sign problem} that diminishes the convergence of Markov chain Monte Carlo calculations, making estimates of observables,
\begin{equation}\label{eq:expectationValue}
  \expval{\hat{O}} =  \frac{1}{\mathcal{Z}}\int \mathcal{D}\phi\,  \hat{O}\left[ \phi \right]e^{-S\left[\phi_n\right]}\approx \frac{1}{N}\sum_{n=0}^N \hat{O}\left[ \phi_n \right]\ ,
\end{equation}
difficult, if not impossible, to obtain.
Reweighting is thus required to handle the otherwise complex probability distribution $e^{-S\left[\phi_n\right]}/\mathcal{Z}$ appearing in every expectation value,
\begin{equation}\label{eq:reweighting}
  \expval{\hat{O}} = \frac{
    \expval{\hat{O} e^{-\mathrm{i}S_I} }_R} {\expval{e^{-\mathrm{i}S_I}}_{R} }
  \approx
  \frac{\sum_{n=0}^N \hat{O}\left[ \phi_n \right] e^{-\mathrm{i}S_I\left[\phi_n\right]}
  }
  {
    \sum_{n=0}^N  e^{-\mathrm{i}S_I\left[\phi_n\right]}
  }\ .
\end{equation}
While including the imaginary phase in the observable and sampling according to the real part of the action is theoretically exact, the statistical convergence is exponentially slowed down by the average phase in the denominator of \autoref{eq:reweighting} due to large phase oscillations. 
We call the absolute value of this phase,
\begin{equation}\label{eq:SP}
  \Sigma\equiv\abs{\expval{e^{-\mathrm{i}S_I}}_{R}}\ ,
\end{equation}
the \emph{statistical power} and use it to quantify the sign problem.  A sign problem-\emph{free} calculation has $\Sigma=1$, whereas the worst sign problem occurs when $\Sigma\sim0$.

Note that for this expectation value the real part of the action is used for the probability distribution, which is a non-holomorphic function.  This means that its value depends on the chosen contour while the physical observables do not. Contour deformation according to Cauchy's theorem is one of many approaches to relieve the sign problem.
We have observed that simple shifts into the complex space can greatly impact the sign problem. It does not have the same capabilities as more intricate transformations such as neural networks, but is simple to apply, does not affect the Hybrid Monte Carlo (HMC) and requires no Jacobian determinant which might blow up the volume scaling \cite{Rodekamp2022, Wynen2020, Ulybyshev:2019fte}. For these reasons we investigate the limits of constant imaginary shifts and develop methods for choosing one.

%To distinguish this proceeding from our paper we also have a look at different lattices, the tetrahedon and the 72-site honeycomb lattice.
\section{Advanced Offsets}
When talking about an offset we are referring to an imaginary constant that is added to all components of the initially real auxiliary field $\phi\to\phi+i\phi_c$.
The most basic, but also most reliable of those offsets occurs at the \emph{tangent plane}. As the name suggests, it is the plane that is tangent to the critical point of the main Lefschetz thimble \cite{Alexandru2016}. It is found by following holomorphic flow equations starting from $\phi=0$ or solving the transcendental equation \autoref{eq:TP}, which we derived for the Hubbard model in~\cite{Gaentgen2023}, at finite imaginary time discretisation $\delta = \beta/N_t$,
\begin{equation}\label{eq:TP}
  \phi_0/\delta = -\frac{U}{N_x}\sum_{k}\tanh\left(\frac{\beta}{2}\left[\epsilon_k+\mu+\phi_0 /\delta\right]\right)\ .
\end{equation}
From this equation we see that the tangent plane scales inversely with the number of time steps $N_t$ and is restricted to the range $[-\tilde{U},+\tilde{U}]$ converging to its boundaries in the limit of $\mu\to\pm\infty$.
We find the tangent plane consistently has a milder sign problem than the unmodified \emph{real plane} calculation. Furthermore we proved that the sign problem vanishes on the tangent plane for $\mu\beta\to\pm\infty$ when $\mu$ is beyond a certain value.

The tangent plane is already known in the lattice community, but varying the offset quickly reveals that even greater improvements can be achieved. 
The next step is a next to leading order correction (NLO) that takes into account thermal fluctuations in the field $\phi$. For this we expand the action around the main critical point and formulate an effective action \autoref{eq:NLO}~\cite{Gaentgen2023}
\begin{equation}\label{eq:NLO}
  S_{\rm eff}[\phi_{1}]=S[\phi_{1}]+\frac{1}{2}\log\det\mathbb H_{S[\phi_{1}]}
\end{equation}
which can be minimized to find $\phi_1$. This offset leads to a further reduction in the sign problem most of the time.%, but is much worse than the tangent plane for usually one small range of $\mu$. To counteract this flaw one could do a test run with both tangent plane and NLO before a larger measurement. 
%The statement "usually one small range of $\mu$" is very confusing, so I deleted it.  T.L.

A further reduction in the sign problem cannot rely on (quasi-)analytical methods alone and requires numerical iterative calculations. Fortunately the derivatives of the statistical power can be easily computed from an existing Markov chain. The derivatives derived in our paper are correct, but \cite{Alexandru:2018fqp} provides a more elegant solution excluding terms that vanish analytically. 
\begin{align}\label{eq:dSP}
	\derivative{}{\phi_0}\expval{e^{-iS_{I,\phi_0}}}_{R,\phi_0} =& \expval{e^{-iS_{I,{\phi_0}}}}_{R,{\phi_0}}\expval{\derivative{S_{R,{\phi_0}}}{{\phi_0}}}_{R,{\phi_0}}
	\\
\label{eq:ddSP}
	\derivative{^2}{{\phi_0}^2}\expval{e^{-iS_{I,{\phi_0}}}}_{R,{\phi_0}} =& \expval{e^{-iS_{I,{\phi_0}}}}_{R,{\phi_0}}\left( 2\expval{\derivative{S_{R,{\phi_0}}}{{\phi_0}}}_{R,{\phi_0}}^2+\expval{\derivative[2]{S_{R,{\phi_0}}}{{\phi_0}}-\derivative{S_{R,{\phi_0}}}{{\phi_0}}^2}_{R,{\phi_0}}\right)
\end{align}
With those and a reasonable initial guess from NLO or tangent plane, this optimal offset that maximizes the statistical power can be approached iteratively. 
To estimate this \emph{optimized plane} we simultaneously fit a Gaussian and its first two derivatives to \autoref{eq:SP}, \autoref{eq:dSP} and \autoref{eq:ddSP} for all previously measured offsets.
We have no formal proof for the statistical power forming a Gaussian around one offset, but judging from our observations it is a sufficiently accurate model for this optimization. This method converges more reliably and slightly quicker than the Newton-Rhapson routine we describe in \cite{Gaentgen2023}.
\begin{figure}
	\centering
	\includegraphics[width=0.7\linewidth]{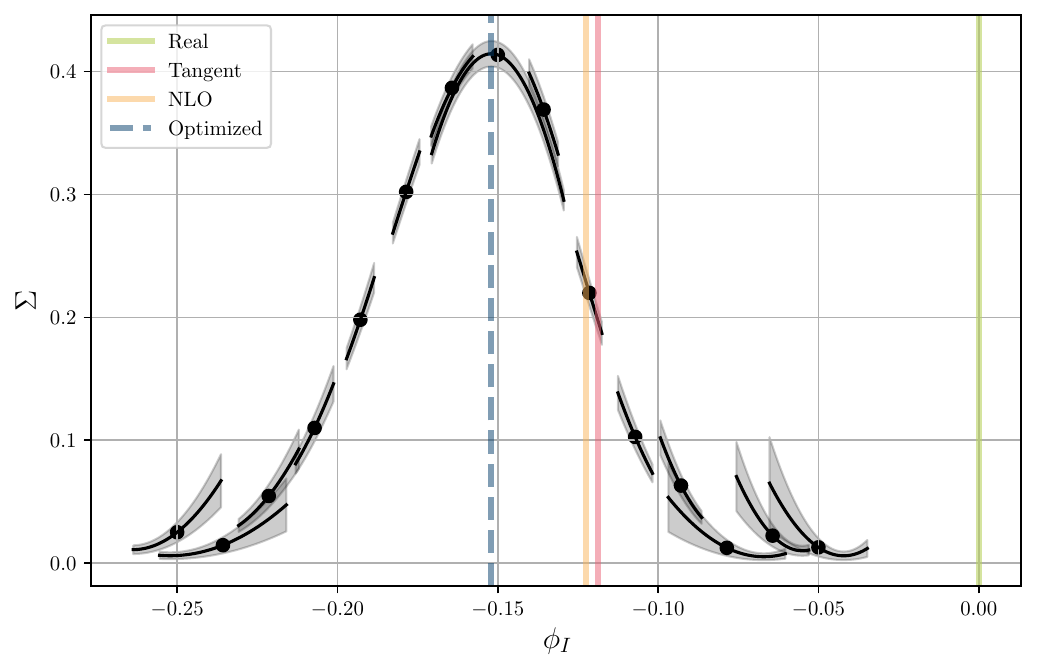}
	\caption{Statistical power with Taylor expansion to second order depending on imaginary offset $\phi_I$. This is the $C_{20}$ fullerene with $N_t=16$, $\beta=6$, $U=2$ and $\mu=1$. The vertical lines mark the named offsets in this example.}
	\label{fig:offsetsp}
\end{figure}
\autoref{fig:offsetsp} illustrates the statistical power with derivatives for a range of offsets showing a clear peak structure motivating this research. The difference in statistical power between the marked offsets emphasizes the importance of the chosen contour. As the effective number of configurations scales with $\Sigma^2$ these offsets are orders of magnitude apart when it comes stochastic uncertainty.

\section{Results}
In this section we apply the developed techniques to a tetrahedron, $C_{20}$, $C_{60}$ and a 72-site honeycomb lattice.
The tetrahedron is small enough for exact diagonalization, allowing us to compare our stochastic algorithm analytic results.

\begin{figure}[h]
	\centering
	\begin{subfigure}[t]{0.48\linewidth}
		\includegraphics[width=\linewidth]{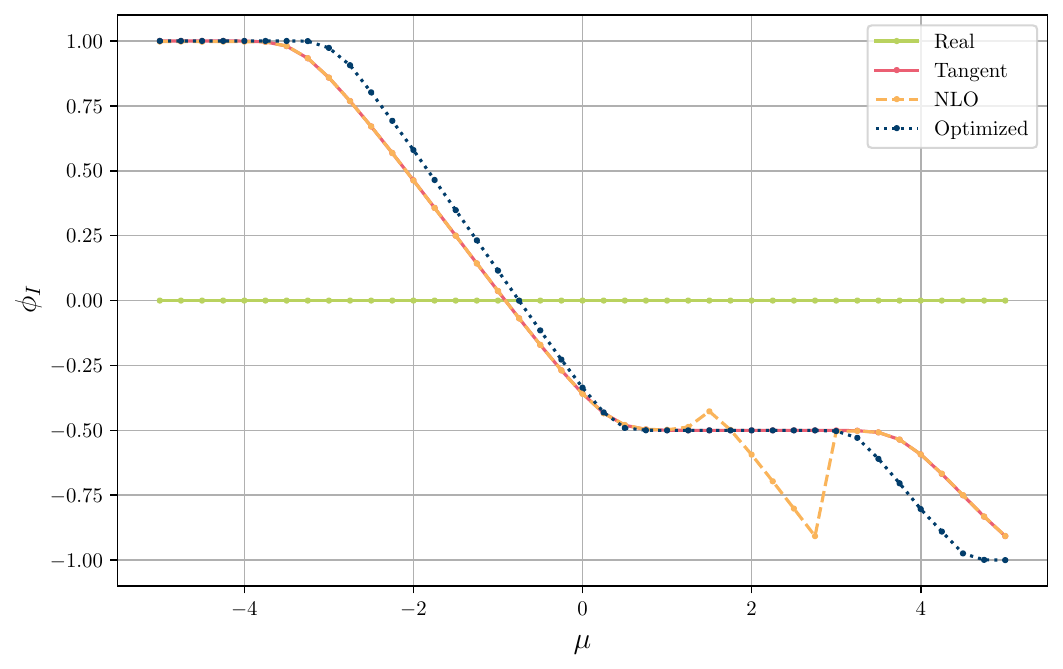}
		\label{fig:offsets}
	\end{subfigure}
	\begin{subfigure}[t]{0.48\linewidth}
		\includegraphics[width=\linewidth]{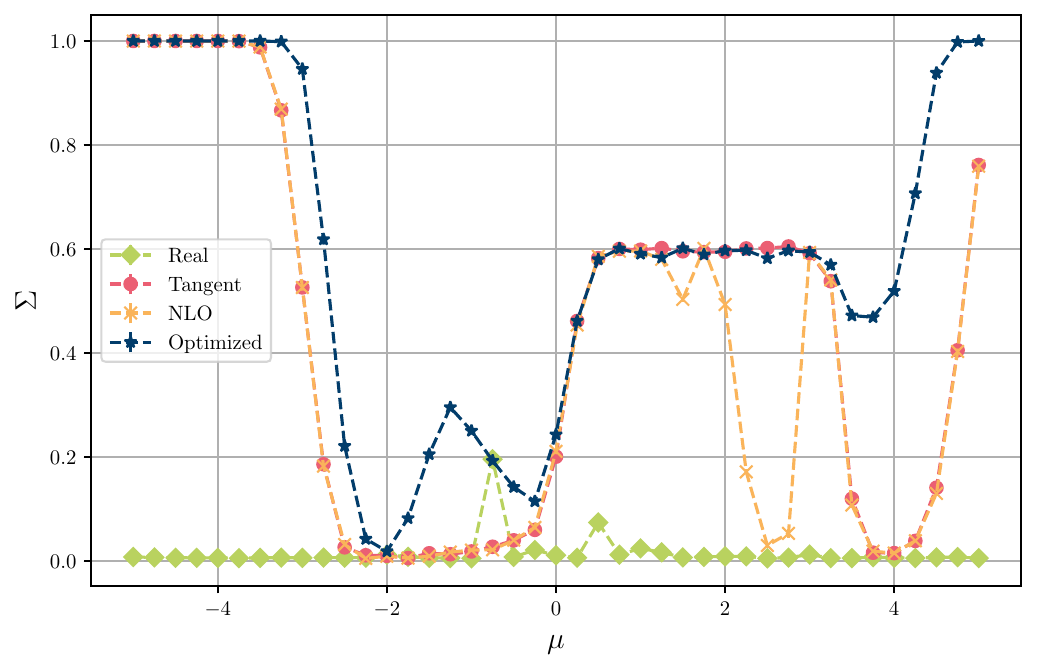}
	\end{subfigure}
	\caption{Comparing the introduced offsets and their corresponding average phase for the tetrahedron with $N_t=16$, $\beta=8$, $U=2$.}
	\label{fig:musp8}
\end{figure}

First, the introduced offsets are quantitatively compared with each other. \autoref{fig:musp8} shows side by side the chosen shift at each $\mu$ and the resulting statistical power.
It confirms our previous statement about the sign problem vanishing for $\mu\to\infty$. But in the interesting range we see that the improvement is greatly dependent on the parameters. At some $\mu$ the tangent plane is already the best we can find, at other $\mu$ the optimized plane surpasses it by magnitudes.
The sign problem also gets worse for increasing $U$ and $\beta$, for more information we refer to \cite{Gaentgen2023}. The sign problem also scales with volume, to compensate this effect we calculate larger lattices at smaller $\beta$~\cite{Splittorff:2006fu}.

\begin{figure}[h]
	\centering
	\begin{subfigure}[h]{0.48\linewidth}
		\includegraphics[width=\linewidth]{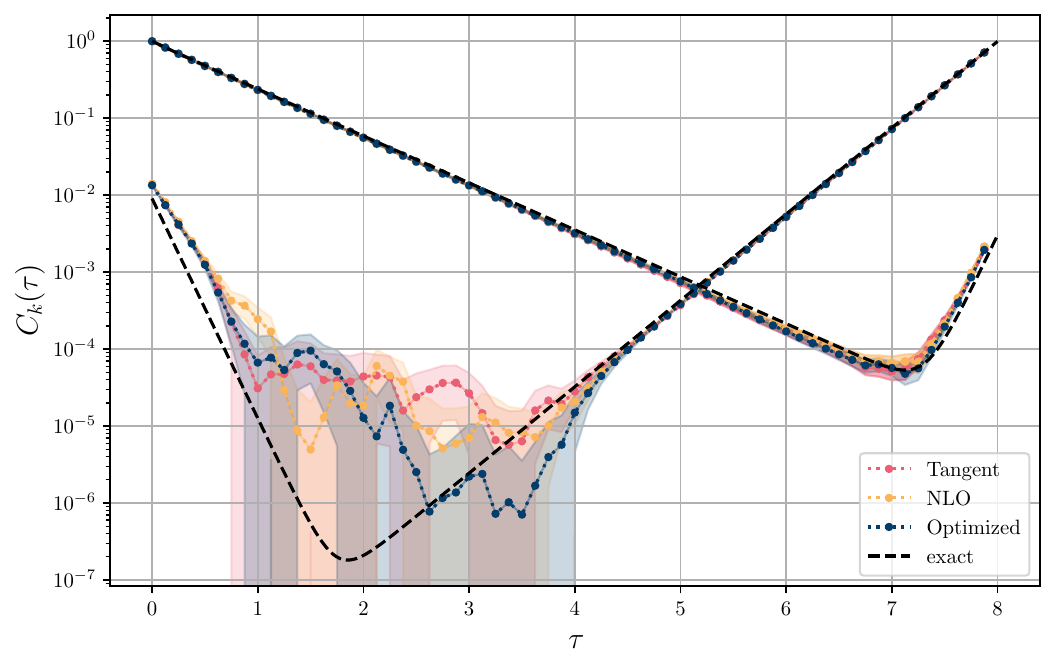}
	\end{subfigure}
	\begin{subfigure}[h]{0.48\linewidth}
		\includegraphics[width=\linewidth]{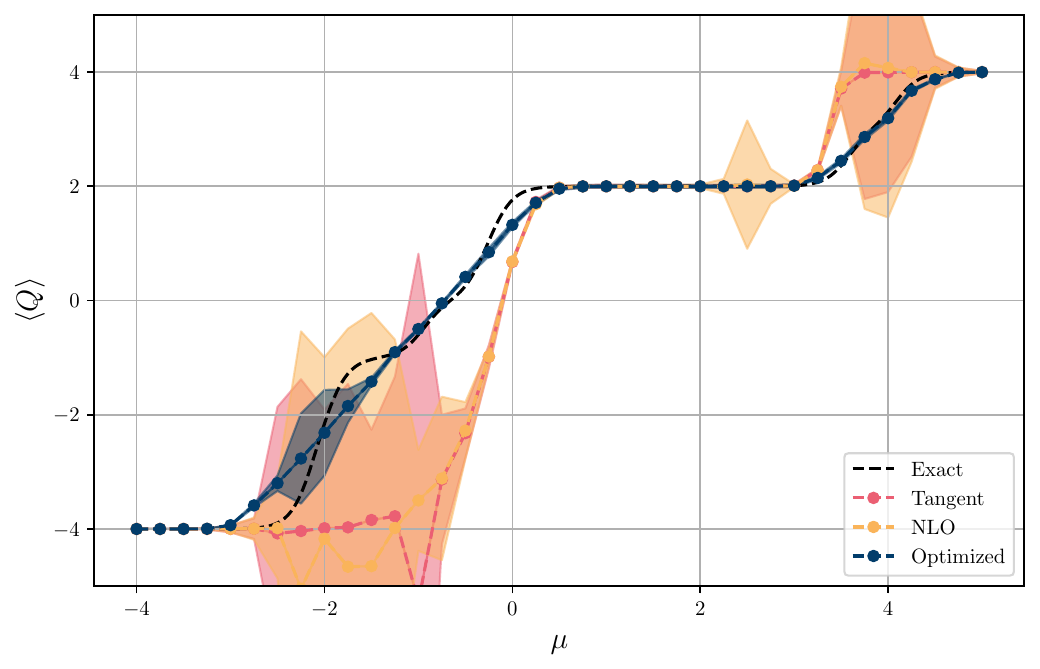}
	\end{subfigure}
	\caption{Observables of tetrahedron with $\beta=8$, $U=2$. Left: single particle correlator $C_k(\tau)$ at $\mu=1$. Right: charge $Q$.}
	\label{fig:restetrahedron}
\end{figure}
\begin{figure}[h]
	\centering
	\begin{subfigure}[h]{0.48\linewidth}
		\includegraphics[width=\linewidth]{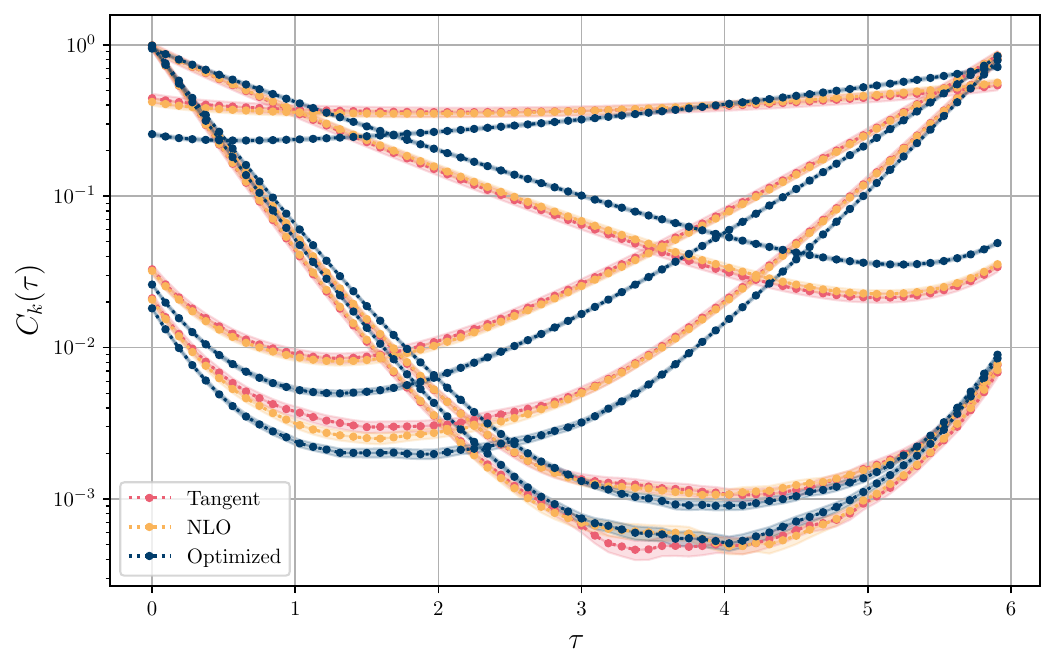}
	\end{subfigure}
	\begin{subfigure}[h]{0.48\linewidth}
		\includegraphics[width=\linewidth]{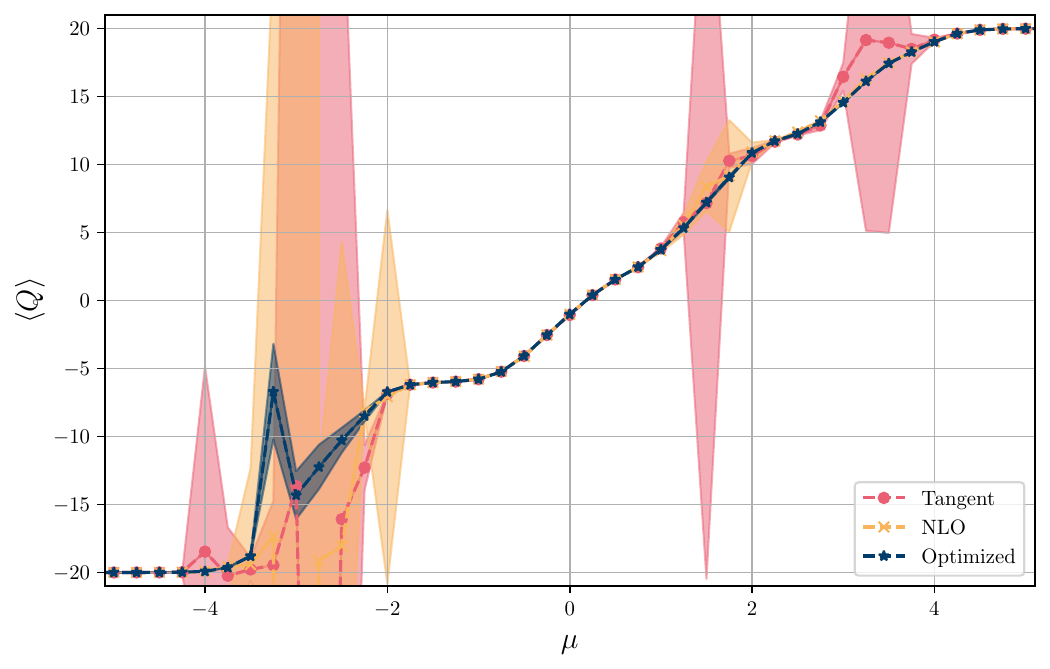}
	\end{subfigure}
	\caption{Observables as in fig.~\ref{fig:restetrahedron} for $C_{20}$ with $\beta=6$, $U=2$ and $\mu=1$.}
	\label{fig:res20}
\end{figure}
\begin{figure}[h]
	\centering
	\begin{subfigure}[h]{0.48\linewidth}
		\includegraphics[width=\linewidth]{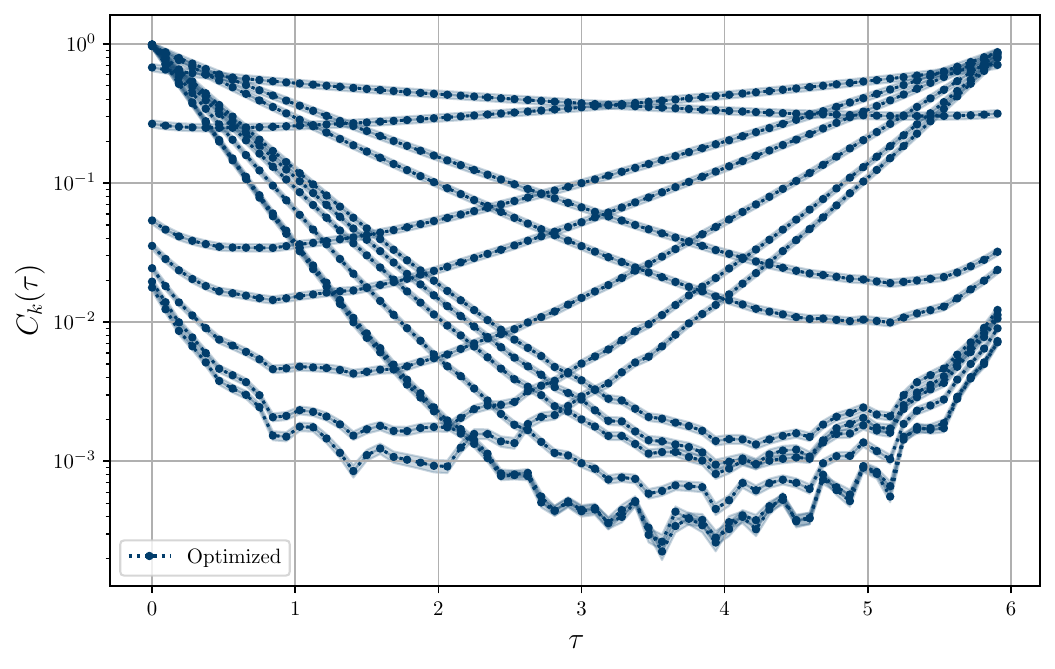}
	\end{subfigure}
	\begin{subfigure}[h]{0.48\linewidth}
		\includegraphics[width=\linewidth]{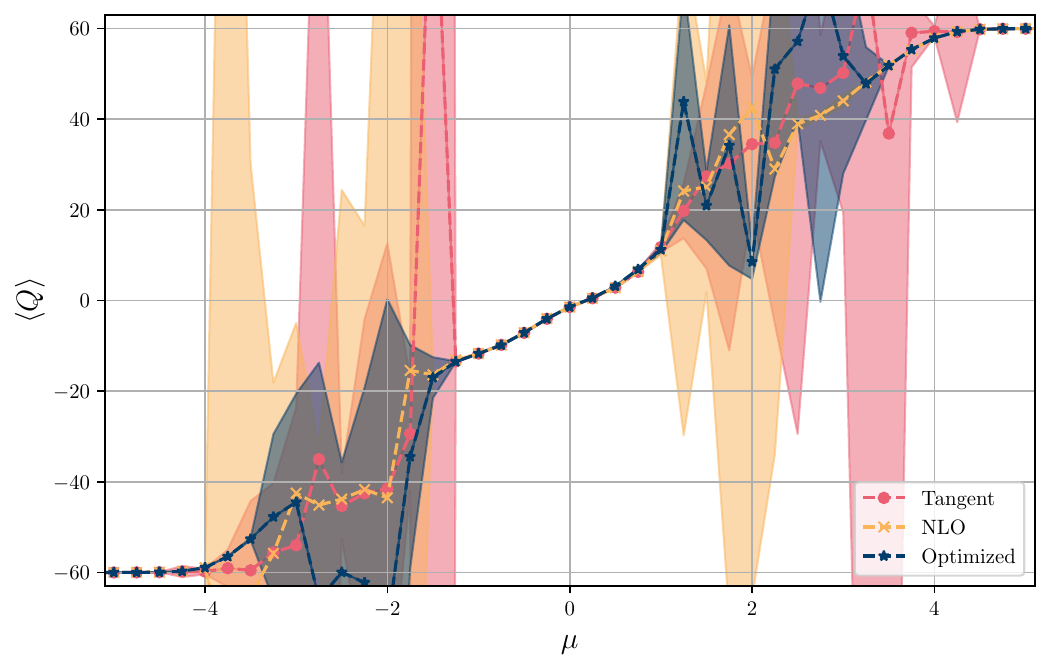}
	\end{subfigure}
	\caption{Observables as in fig.~\ref{fig:restetrahedron} for $C_{60}$ with $\beta=6$, $U=2$ and $\mu=1$.}
	\label{fig:res60}
\end{figure}
\begin{figure}[h]
	\centering
	\begin{subfigure}[h]{0.48\linewidth}
		\includegraphics[width=\linewidth]{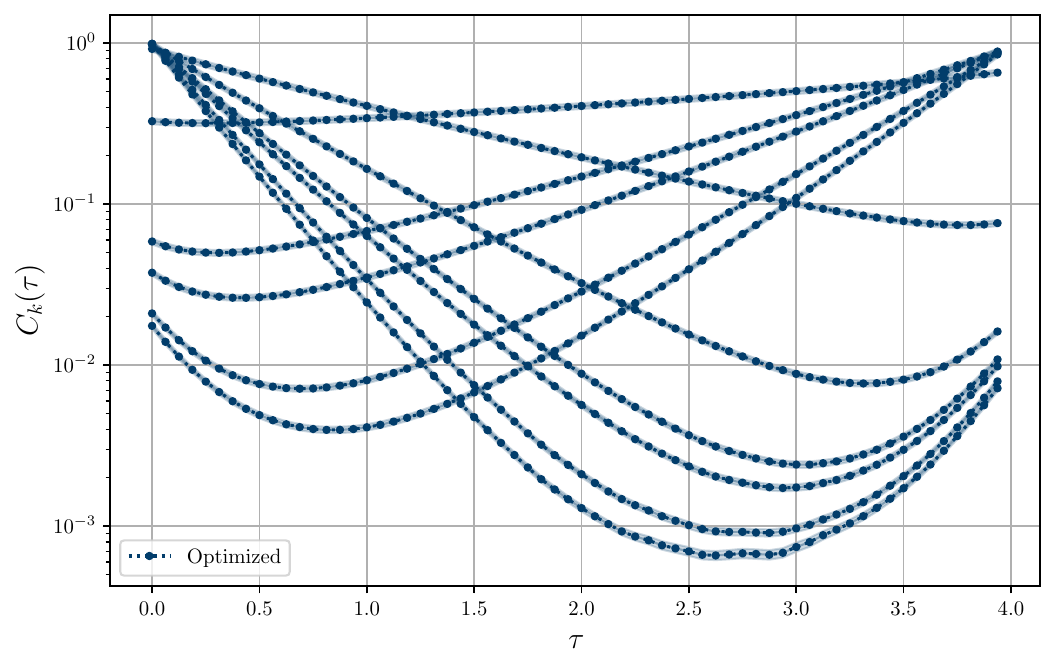}
	\end{subfigure}
	\begin{subfigure}[h]{0.48\linewidth}
		\includegraphics[width=\linewidth]{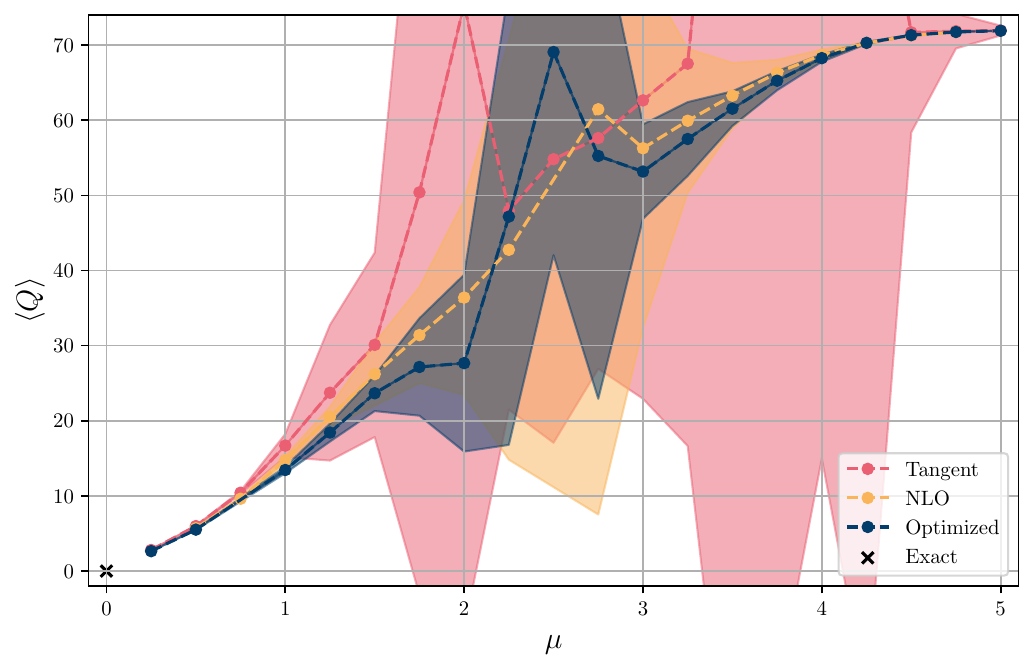}
	\end{subfigure}
	\caption{Observables as in fig.~\ref{fig:restetrahedron} for 72-site honeycomb lattice ($6\times6$ unit cells) with $\beta=4$, $U=2$ and $\mu=1$.}
	\label{fig:res72}
\end{figure}

Furthermore, we show exemplary observables to highlight the quality difference that the choice of contour can cause even in this most basic form. 
The first observable is the single particle
\footnote{ 
    Here we use the Hubbard model in particle-hole basis where a particle are annihilated with 
    $a_x = a_{x,\uparrow}$ and holes with $b_x = a^\dagger_{x,\downarrow}$. The creation operators are similar.
}
correlation function given by \autoref{eq:spc}. 
It can be used to extract the low lying spectrum relative to the ground state, because at large $\beta$ the smallest transition energies will dominate the slope at $\tau\to0$ or $\tau\to\beta$:
\begin{equation}\label{eq:spc}
	C_k(\tau) = \expval{a_k^{}(\tau)a_k^\dagger(0)} = \sum_{ a,b}\abs{\bra{\Psi_a}a_k^\dagger\ket{\Psi_b}}^2e^{-\tau (E_b- E_a)}e^{-\beta E_a}
\end{equation}
The k-index labels the momenta rather the positions on the spatial lattice. For more details consider~\cite{Ostmeyer:2020uov}. The systems we work with have degenerate energy states. We average the corresponding correlation functions and show in our figures only the resulting selection of unique momenta.

The second observable is the total charge expectation value \autoref{eq:charge} which is an experimentally measurable quantity, depending on the chemical potential $\mu$,
\begin{equation}\label{eq:charge}
	\expval{Q} = \expval{\sum_x q_x } = N_x - 2\sum_k C_k(\tau=0)\ .
\end{equation}
Comparing these charge expectation values with the ones from \cite{Gaentgen2023} shows that the afore mentioned Gaussian fits improved the stability of our optimization routine in areas where the sign problem is especially strong.
\subsection*{Energy Fits}
It is clear from \autoref{eq:spc} that the ground state has the greatest contribution, especially at large $\beta$. We fit an exponential to each correlation function shown in figures \ref{fig:restetrahedron}, \ref{fig:res20}, \ref{fig:res60} and \ref{fig:res72} to determine the particle excitation energy.

\begin{figure}[h]
	\centering
	\begin{subfigure}[h]{0.48\linewidth}
		\includegraphics[width=\linewidth]{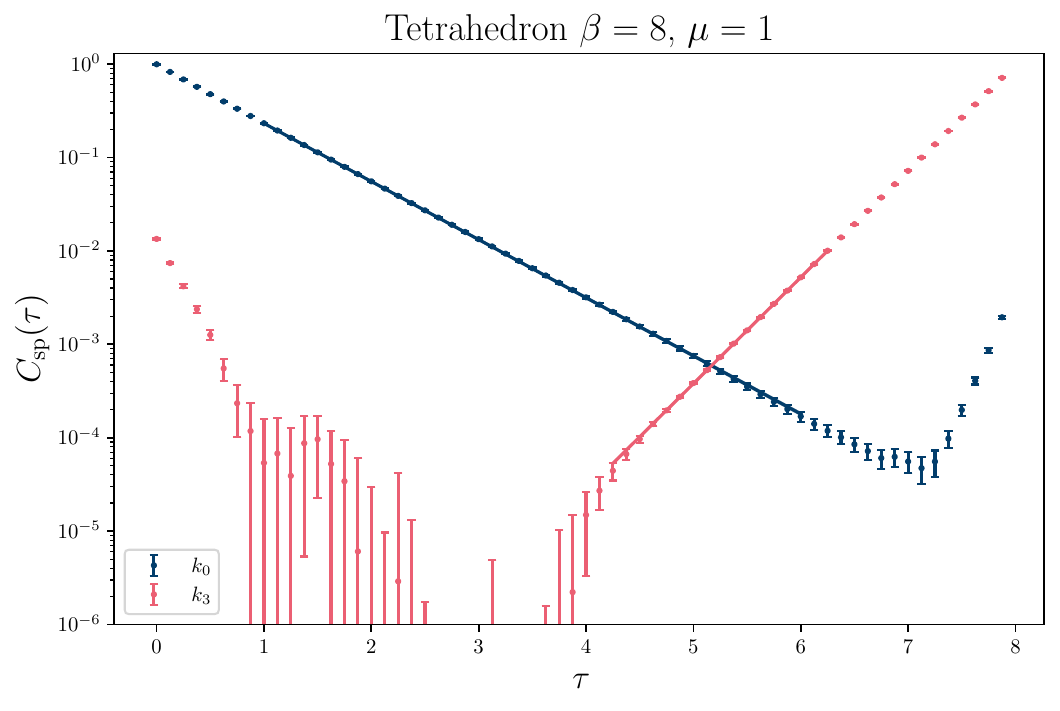}
	\end{subfigure}
	\begin{subfigure}[h]{0.48\linewidth}
		\includegraphics[width=\linewidth]{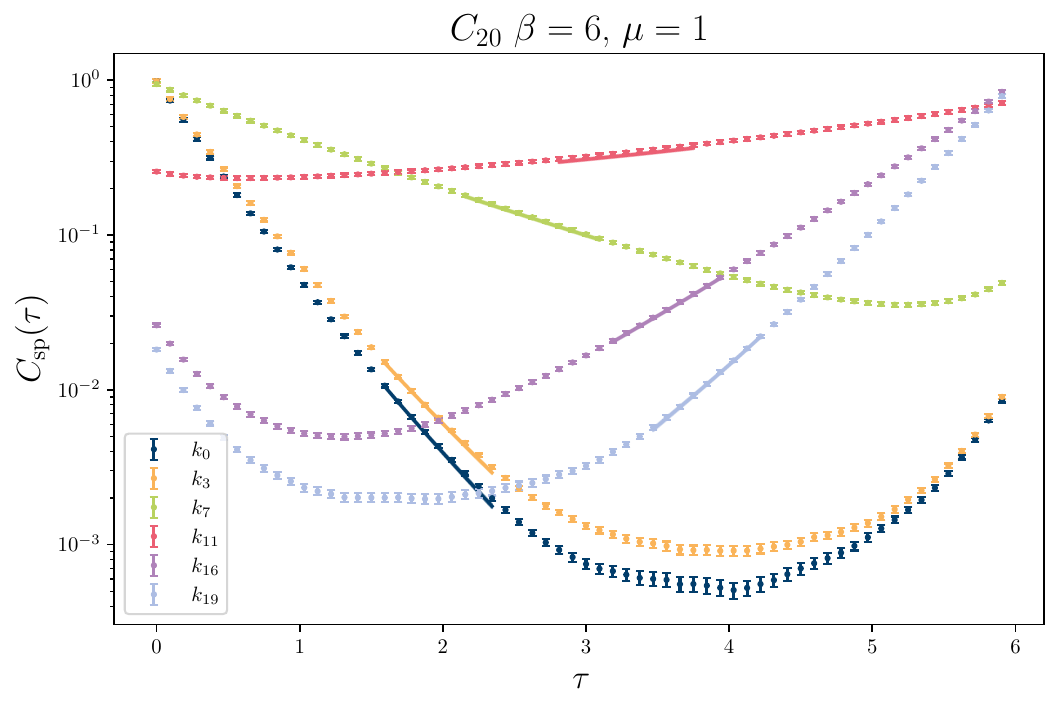}
	\end{subfigure}
	\caption{Single particle correlation functions with exponential fit. Both lattices have $U=2$ and $N_t=64$.}
	\label{fig:fits}
\end{figure}

\begin{figure}[h]
	\centering
	\begin{subfigure}[t]{0.49\linewidth}
		\centering
		\includegraphics[width=1\linewidth]{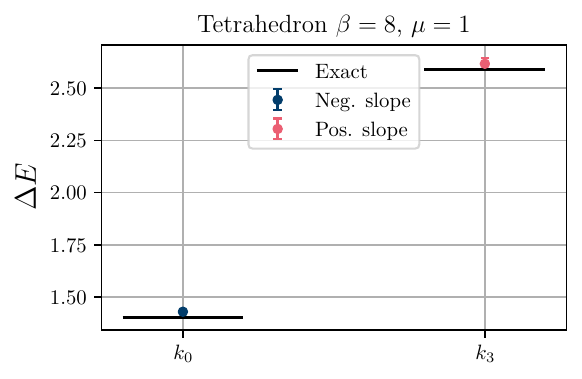}
	\end{subfigure}
	\begin{subfigure}[t]{0.49\linewidth}
		\centering
		\includegraphics[width=1\linewidth]{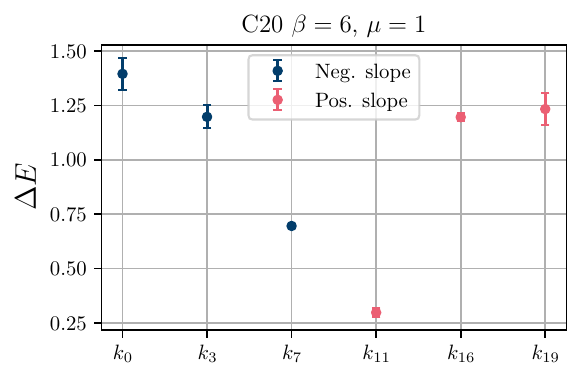}
	\end{subfigure}
	\begin{subfigure}[t]{0.49\linewidth}
		\centering
		\includegraphics[width=1\linewidth]{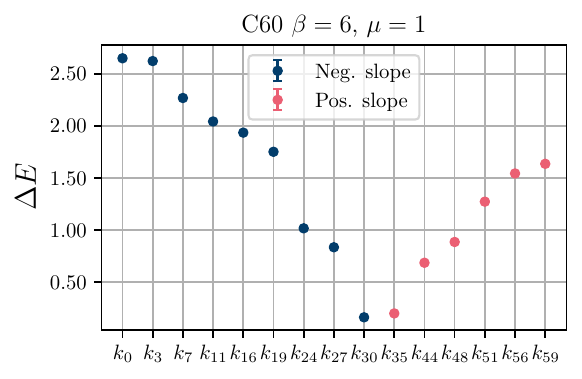}
	\end{subfigure}
	\begin{subfigure}[t]{0.49\linewidth}
		\centering
		\includegraphics[width=1\linewidth]{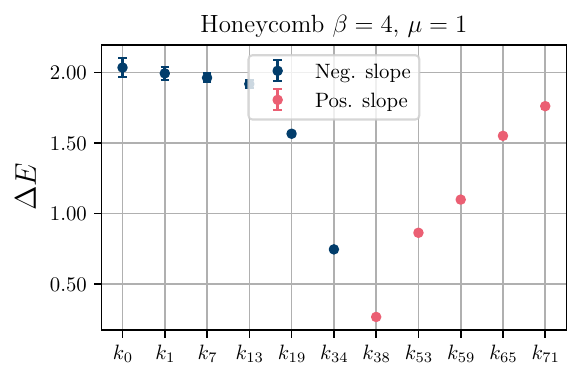}
	\end{subfigure}
	\caption{Energy states reachable by particle creation/annihilation from ground state. All systems have $U=2$ and $N_t=64$.}
	\label{fig:energies}
\end{figure}

\autoref{fig:fits} illustrates our fits for tetrahedron and $C_{20}$.
We identify negative slopes (i.e.\ positive energies) as $\Delta N=+1$ transitions and positive slopes as $\Delta N=-1$ transitions, i.e. when the ground state has particle number $N$ the excited state associated with a single particle correlation function has particle number $N+\Delta N$. In \autoref{fig:energies} we show the resulting energies for all four lattices. Again we provide exact values for the tetrahedron that we calculated from exact diagonalization accounting for discretized time. Keep in mind that the spectrum is shifted due to the chemical potential and the ground state might not be at half filling.

%\subsection*{Statistical Power and the Charge}
%Maybe I'll add our observation with the charge here
%\begin{figure}[h]
%	\centering
%	\includegraphics[width=0.7\linewidth]{figure/meff0_5}
%	\caption{}
%	\label{fig:meff05}
%\end{figure}
\newpage
\section{Summary \& Outlook}
We demonstrate that a smart choice of the integration contour can drastically reduce the sign problem in the Hubbard model without compromising numerical efficacy. Even a basic transformation like a constant shift can improve the quality of HMC calculations by orders of magnitude while entailing only negligible computational costs and small human effort. These methods cannot solve the exponentially hard sign problem, but they can expand the explorable parameter space at a given budget.

We present our methods of finding such a favorable offset and compare them with each other. All of them perform much better than the default calculation on the real plane.
The recently developed optimization routine, utilizing fits to the statistical power and its derivatives, provides more stability than the initial version relying on the Newton-Rhapson method. 
Also we show the extraction of energy levels from the correlation functions and confirm that they align with the exact results on a small system.

In the future we intend to apply these methods to more carbon nano-systems to provide a detailed analysis of their electronic properties. We will also continue the exploration of simple contour deformations that do not induce a Jacobian with limiting volume scaling. Some current ideas that come into question are individual offsets for each component of $\phi$ and offsets combined with a Gaussian around $0$. We hope that these investigations will deepen our understanding of sign optimizing manifolds and provide better starting points for the training of neural networks.

\newpage
\section*{Acknowledgments}
We thank Neill Warrington for many helpful discussions related to this work
as well as Timo Lähde for his valuable comments.
This work was funded in part by the
Deutsche Forschungsgemeinschaft (DFG, German Research Foundation) through the funds provided to the
Sino-German Collaborative Research Center “Symmetries and the Emergence of Structure in QCD” (NSFC
Grant No. 12070131001, DFG Project-ID 196253076
– TRR110) as well as the STFC Consolidated Grant
ST/T000988/1.
This work is supported by the MKW NRW under the funding code NW21-024-A.
We gratefully acknowledge the computing time
granted by the JARA Vergabegremium and provided on
the JARA Partition part of the supercomputer JURECA
at Forschungszentrum Jülich.
\bibliographystyle{abbrv}
\bibliography{references.bib}
%\begin{thebibliography}{99}
%\bibitem{...}
%....

%\end{thebibliography}

\end{document}